\documentclass[twocolumn,aps,pra,showpacs]{revtex4-1}
\usepackage{graphicx}
\usepackage{graphics}
\usepackage{color}

\newcommand{\nn}{\nonumber}

\begin{document}
\def\tilx{{\tilde{x}}}
\def\tilph{{\tilde{\phi}}}
\def\tilE{{\tilde{\cal{E}}}}
\def\tilz{\tilde{z}}
\def\d{{ d}}

\title{Instanton theory for bosons in disordered speckle potential}

\author{G. M. Falco$^1$, and  Andrei A. Fedorenko$^2$}
\affiliation{ $^1$Institut f\"ur Theoretische Physik, Universit\"at zu
K\"oln, Z\"ulpicher Str. 77, D-50937 K\"oln, Germany \\
$^2$CNRS UMR5672 -- Laboratoire de Physique de l'Ecole Normale
Sup{\'e}rieure de Lyon, 46, All{\'e}e d'Italie, 69007 Lyon, France }

%\date{\today}
\date{May 7, 2015}

\begin{abstract}

We study the tail of the spectrum for non-interacting bosons in
a blue-detuned random speckle potential.
Using an instanton approach
we derive the asymptotic behavior of the density of states
in $d$ dimensions. The leading corrections resulting from fluctuations around
the saddle point solution are obtained
by means of the Gel'fand-Yaglom method generalized to functional
determinants with zero modes. We find a good agreement with the results
of numerical simulations in one dimension. The effect of weak repulsive
interactions in the Lifshitz tail is also discussed.

\end{abstract}

\maketitle
%%%%%%%%%%%%%%%%%%%%%%%%%%%%%%%%%%%%%%%%%%%%%%%%%%%%%%%%%%%%%%%%%%%%%%%%%%%%%%%%

\section{Introduction}

Effect of disorder on quantum systems has been attracted considerable
interest in condensed matter physics during the last several decades.
In recent years, it was realized that ultracold atomic gases in optical
speckle potentials may serve as quantum simulators for diverse
phenomena in disordered quantum systems~\cite{GModugno10,Palencia10,Shapiro12}.
Contrary to the experiments in condensed matter physics,
the optical speckles allow one to create a controllable random potentials
acting on ultracold atoms~\cite{Shapiro12,Boiron99}. Many interesting  features
of Bose-Einstein condensates (BECs) in disordered  speckle potentials have
been addressed from both the experimental and the theoretical
sides~\cite{Lye05,Clement05,Fort05,Schulte05,Palencia06}.
These include inhibition of transport properties~\cite{Clement05,Fort05,Clement06},
fragmentation effects~\cite{Lye05,Yong08,Dries10},
frequency shifts~\cite{Lye05,Modugno06},
damping of collective excitations~\cite{Lye05,Dries10,Modugno06} and
Anderson localization~\cite{Palencia07,Lugan07,Billy08,Roati08,Palencia08,%
Kondov11,Gehee13,Giacomelli14}. Also, the superfluid-insulator
transition~\cite{Fisher89,Zhou06,Diallo07,Falco09,Gurarie09,Fontanesi10,Altman10,%
Bissbort10,Deissler10} and the transport of coherent matter waves have been recently
investigated from the theoretical point of view for speckles in
higher dimensions~\cite{Kuhn05,Kuhn07,Hensler08,Fallani08,Pilati10,Orso14}.

When a coherent laser light is scattered from a rough surface the
partial waves passing through the  different parts of the surface acquire
random phase shifts.  The interference of these randomly phased waves
produces a speckle pattern which consists of the regions or grains
of light intensity with random magnitude, size and position.
The local intensity of the speckles, $I (x) = |{\cal{E}}(x)|^2$,
is determined by the electric field ${\cal E}(x)$. To a very good
approximation, the electric field ${\cal E}(x)$ can be viewed as a complex
Gaussian variable with finite correlation
length $\xi$ giving the typical size of the light intensity grains in the speckle
pattern.
The distribution of intensity $I (x)$ across the
speckle pattern follows a negative-exponential (or Rayleigh)
law~\cite{Modugno06,Huntley89},
\begin{equation}
\label{eq:spkldistr}
P\left[I\right] = \exp{\left[-I/I_0\right]/I_0},
\end{equation}
where $I_0 = \left\langle I \right\rangle  $ is the mean intensity while
the most probable intensity is zero.
The speckle pattern shined on a sample
of atoms creates a random potential felt by the
atoms provided that the wavelength of the laser light is slightly detuned
from the atomic resonance.
The potential is proportional to the local light intensity,
$V (x) = \alpha~I(x)$, so that
the non-interacting atoms in the speckle potential can be described by the
Schr{\"o}dinger equation
\begin{eqnarray}
\left[-\frac{\hbar^2}{2m_0}\nabla^2 + \alpha ~I\left(x\right)
\right]\psi\left(x\right)= E~ \psi\left(x\right),
\end{eqnarray}
where $m_0$ is the mass of atoms and  $\psi\left(x\right)$ is
the single particle wave function.
The constant $\alpha$ is proportional to the inverse of the
detuning $\Delta$ between the laser and the atomic resonance. The detuning
can be either positive (blue detuning) or negative
(red detuning)~\cite{Fallani08}.
The blue-detuned case corresponds to a disordered potential
consisting of a series of barriers bounded from below. The red-detuned speckle
produces a potential bounded from above and made of potential wells.

The precise single particle spectrum of the speckle potential
is unknown even in the 1D case despite the intense research activity in this field.
It is widely believed that
the density of states (DOS) of blue-detuned repulsive speckles is
characterized by a usual Lifshitz tail for potentials bounded from
below~\cite{Lifshitz88}.
However, many exact results known for 1D random
potentials~\cite{Lifshitz88,Kramer93} cannot be directly applied to
the speckle potential since it is correlated and non-Gaussian.
In the previous paper~\cite{Giacomelli10} we investigated analytically
and numerically  the 1D single-particle
spectrum for the both red-detuned and blue-detuned speckle potentials.
Since the speckle pattern is characterized, besides
the mean intensity $I_0$, by the correlation length $\xi$, this
introduces  a new energy scale $E_{\xi}=\hbar^2 /2 m_0 \xi^2$.
We have shown that for dimensional reasons,
the single-particle properties are determined by the dimensionless parameter
$s =  2 m_0 \xi^2 \alpha I_0/\hbar^2$ for the both red and blue detuned speckle
potentials.
We identified different Lifshitz regimes controlled by the dimensionless parameter
$s$ which vary from the semiclassical limit $|s| \gg 1 $ deep to the quantum
limit $|s| \ll 1$~\cite{Giacomelli10}.

In the present paper we improve our results for the blue-detuned speckle in 1D
and extend them to higher dimensions.
We apply the field-theoretic description developed by Luck and
Nieuwenhuizen~\cite{Luck89}  for studying the Lifshitz tails in the random potentials
bounded from below.
We generalize  this approach to the blue-detuned speckle potential which is not
only bounded from below but also has a
finite correlation length.
We argue that well below the corresponding energy scale $E_{\xi}$
the Lifshitz tail does not depend on
the precise form of the electric field correlations since the DOS is determined
by the states localized in the regions with very small intensity whose size is much
larger than $\xi$. In this regime the instanton approach proposed
in Ref.~\cite{Luck89} for the potential bounded from below
predicts the following form of the DOS tail
\begin{eqnarray}\label{eq:dens-saddle-Luck}
\nu\left(E\right)= {\cal A}_d(E)~\exp{\left[-v_{\rm d}
~F_d\left[\ln{\left({I_0}/{E}\right)}\right]
\left({E_{\xi}}/{E}\right)^{d/2}\right]},\ \ \
\end{eqnarray}
where $v_{\rm d}$ is a constant which depends on the dimension and
$F_d$ and ${\cal A}_d$ are some functions of the energy.

The paper is organized as follows. In Sec.~\ref{sec:action} we derive the
replicated action for a particle in a blue-detuned speckle potential.
The saddle point solution of the classical
equation in $d$ dimensions is discussed in Sec.~\ref{sec:instanton} where
the expressions for $v_{\rm d}$ and $F_d$ are calculated.
The fluctuations around the instanton solution are investigated in
Sec.~\ref{sec:fluct}.
There, the prefactor ${\cal A}_d $ is calculated  using
the Gel'fand-Yaglom (GY) method~\cite{Gelfand60,Kleinert04}
generalized  by Tarlie and McKane (TM)~\cite{Tarlie95} to
functional determinants with excluded zero modes. In order to illustrate
the power of this method, in Appendix~\ref{appendix-C} we reconsider
the problem of a particle in Gaussian uncorrelated disorder for which
there are some discrepancies in the existing literature.
In Sec.~\ref{sec:interactions} we consider a weakly  interacting
Bose gas in a speckle potential. The obtained results are summarized in
Conclusion.

\section{Averaging over disorder: replicated action }
\label{sec:action}
The DOS for a particle in a particular realization of the electric
field ${\cal E}(x)$ can be related to
the imaginary part of the one-particle Green function
\begin{eqnarray}\label{eq:dens}
\nu(E)= -\frac{1}{\pi} \textrm{Im}\ {{\cal G}(x,x;E)}.
\end{eqnarray}
To average over different realizations of disorder potential we employ
the replica trick~\cite{Cardy78}. We introduce $N$  replicas of the
original system and use the functional integral representation
\begin{eqnarray}\label{eq:func-int}
{\cal G}(x,x';E) ={\lim_{N\rightarrow 0}} \int \mathcal{D}\phi
~\phi_1(x) \phi_1(x')~
 e^{-S[\bar{\phi}]}
\end{eqnarray}
with the action
\begin{eqnarray}\label{eq:func-action}
S[\bar{\phi}]=\frac{1}{2}\int d^{d}x \left\{ \frac{\hbar^2}{2m_0}
\left(\nabla \bar{\phi}(x)\right)^2
+\left[V(x)-E\right]{\bar{\phi}}^{\,2}(x)\right\},\ \ \ \
\end{eqnarray}
where $\bar{\phi}(x)$ is an $N$-component scalar field.
The disorder potential is proportional to the local intensity of the
speckles pattern created by a laser, $V(x)=\alpha~ {\cal E}^*(x) {\cal E}(x)$, where
${\cal E}(x)$  is the electric field and we fix $\alpha=+1$
in the case of a blue-detuned speckle.
To a very good approximation,  the electric field is
a random complex Gaussian field  with zero mean and variance
\begin{equation}
\left \langle{\cal E}^*(x){\cal E}(y) \right\rangle = G(x - y),
\label{eq:G-def1}
\end{equation}
where the function $G(x - y)$ has the width of $\xi$ and its precise form
depends on the experimental setup~\cite{Palencia08,Kuhn07}.
Then the average of
the disorder potential can be expressed as exponential of the sum of loop diagrams:
\begin{eqnarray}
&& \! \! \left\langle \exp[- \int d^{d}x ~ {\cal E}^*(x){\cal E}(x)
\bar{\phi}^2(x) ]
\right\rangle  \ \ \  \nonumber \\
&& \
=\exp \left\{
- \frac{1}{1!}
\parbox{8mm}{\includegraphics[width=8mm]{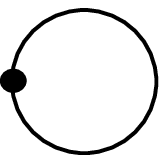}} +
\frac{1}{2!}
\parbox{8mm}{\includegraphics[width=8mm]{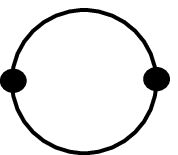}} -
\frac{1}{3!}
\parbox{8mm}{\includegraphics[width=8mm]{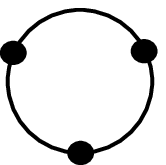}} +...
\right\}. \ \ \label{expEE}
\end{eqnarray}
In Eq.~(\ref{expEE})
the lines  stand for  the correlator~(\ref{eq:G-def1})
and thus have two distinct ends corresponding
to ${\cal E}^*$ and ${\cal E}$.
Two lines can be connected only by ends of different types and
the corresponding vertex  carries a factor of
$\frac12 \bar{\phi}^2(x)=\frac12\sum_{n=1}^N \phi_n^2(x)$.
The loop diagram with $j$ vertices in Eq.~(\ref{expEE}) has the combinatorial
factor of $(j-1)!$ which gives the number of possibilities to construct a
loop from $j$ lines with distinct ends.
Summing up the all diagrams for an arbitrary  function $G(x)$
and field $\bar{\phi}(x)$  is a formidable task.
However, there are several cases  when one can solve this problem at
least partially. For a special class of correlation functions
$G(x)$ the sum of the diagrams can be rewritten as a ratio of functional determinants
(see Appendix~\ref{sec:app:1}).
The summation of the diagrams can be also performed within a variational method
with Gaussian correlators and trial functions (see Appendix~\ref{sec:app:2}).
These approximations can be used to show that the low-energy  tail  of the DOS
does not depend on the precise form of the electric field correlator
for $E\ll E_\xi$ and it is completely determined by $E_\xi$ and $I_0$.
This is in contrast to the Gaussian unbounded potential~\cite{Cardy78} where
the presence of correlations changes the low-energy Lifshitz tail of the
DOS~\cite{John84}.
For $E\ll E_\xi$ we can approximate the electric field correlator by
$$G(x-y)=I_0 \xi^{\d} \delta_{\xi}^{\d}(x-y)$$
where we have defined the regularized $\delta$-function
of width $\xi$ such that
\[
\lim\limits_{\xi\to 0}\delta_{\xi}^{ d}(x-y)=\delta^{ d}(x-y)
\hspace{7mm} \mathrm{ and } \hspace{7mm}
\delta_{\xi}^{ d}(0) = \frac{1}{\xi^d}.
\]
As a result the loop diagram with $j$ insertions of $\bar{\phi}^2(x)$
can be expressed as
\begin{equation}
 \frac{(-1)^j}{j!} \,
\parbox{13mm}{\includegraphics[width=13mm]{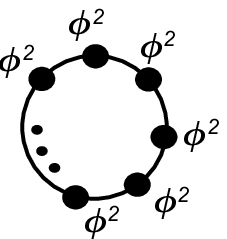}}
~=~\int \frac{d^{d}x}{\xi^d} \frac{(-1)^j}{j 2^j} (I_0 \xi^d)^j \bar{\phi}^{2j}(x).
\end{equation}
By using the relation
\begin{eqnarray}
\sum\limits_{j=1}^{\infty}  \frac{(-1)^j}{j 2^j} (I_0 \xi^d)^j \bar{\phi}^{2j}(x)
= - \ln \left[1+ \frac12(I_0 \xi^d) \bar{\phi}^{2}(x) \right], \nonumber
\end{eqnarray}
we can sum up all the diagrams. The averaged replicated action then reads
\begin{eqnarray}\label{eq:action-av}
S_{\rm av}=\int d^{d}x &\left\{\frac{\hbar^2}{4m_0}(\nabla \bar{\phi}(x))^2
-\frac{1}{2}E\bar{\phi}^2(x)
\right.\nonumber\\
&\left.+ \frac1{\xi^d} \ln \left[1+\frac{1}{2}(I_0 \xi^d) \bar{\phi}^{2}(x)
\right]\right\}.
\end{eqnarray}

%%%%%%%%%%%%%%%%%%%%%%%%%%%%%%%%%%%%%%%%%%%%%%%%%%%%%%%%%%%%%%%%%%%%%%%%%%

\section{Saddle point solution}
\label{sec:instanton}

The field theory~(\ref{eq:action-av}) has a trivial vacuum
state $\bar{\phi}=0$.
However, a perturbative expansion around it
does not contribute to the DOS at any finite order
since the Green function remains real in this approximation.
Following Ref.~\cite{Luck89,Cardy78}
we assume that the functional integral~(\ref{eq:func-int}) is dominated by a
spherically symmetric saddle point field configuration.
The integration over fluctuations around this instanton solution
brings a finite imaginary part
to the Green function ${\cal G}$ leading to a finite
DOS.

It is convenient to express the action~(\ref{eq:action-av}) in terms  of
the rescaled quantities
$\bar{\phi}(x) =  \sqrt{2}\bar{\tilde{\phi}}({x}/{\xi} )/\sqrt{E_{\xi} \xi^d}$,
$\tilde{x}={x}/{\xi}$ and $\tilde{E}=E/E_{\xi}$.
By omitting the tildes on $\bar{\phi}$ and $x$ we arrive at
\begin{eqnarray}
S_{\rm av}=  \int d^d{x} \left\{ \left[\nabla_{{x}}
{\bar\phi}({x})\right]^2 -\tilde{E} {\bar\phi}^2({x})
+ \ln \left[1+ s {\bar{\phi}}^{2}(x)\right]\right\}. \nn \\
\label{eq:Snewresc}
\end{eqnarray}
The variational principle gives the following classical equation of motion
\begin{eqnarray}\label{eq:schroedingernew}
{\nabla}^2 {\bar\phi}+\tilde{E}{\bar\phi}=\frac{{\bar\phi}}{s^{-1}+{\bar\phi}^2}.
\end{eqnarray}
Assuming that Eq.~(\ref{eq:schroedingernew}) has a solution of the form
\begin{equation}\label{eq:phi-clas}
\bar\phi_{\rm cl}(x)=\bar{n}~\phi_0(x)
\end{equation}
with $\bar{n}^2=1$, we rewrite it as
\begin{eqnarray}\label{eq:schroedingerphinew}
{\nabla}^2 \phi_0+\tilde{E}\phi_0=\frac{\phi_0}{s^{-1}+\phi_0^2}.
\end{eqnarray}
It is instructive to compare this equation with
the corresponding saddle point equation~(\ref{eq:schroedingerCardy})
in the case of $\delta$-correlated Gaussian disorder
(see~\cite{Cardy78} and Appendix~\ref{appendix-C}).
At variance with the Gaussian disorder,
one cannot eliminate the
explicit energy dependence in Eq.~(\ref{eq:schroedingerphinew}) by any
variable transformation.
In the limit $E\rightarrow 0$,
the classical solution of Eq.~(\ref{eq:schroedingerphinew}) approaches the
form
\begin{eqnarray}\label{eq:newresc-f}
\phi_0(x)\approx \sqrt{{a_d \ln{\left({s}/\tilde{E}\right)}}/{\tilde{E}}}~L_d\left(
x \sqrt{\tilde{E}}\right)
\end{eqnarray}
in the region  $0\leq x \sqrt{\tilde{E}} \leq \mu_d$ and essentially vanishes
elsewhere.
The functions $L_1(t)=\cos{t}$, $L_2(t)=J_0(t)$ and $L_3(t)=\sin{t}/t$ are
the spherical Bessel functions in $d$ dimensions,
$\mu_d$ is the first zero of $L_d$ ($\mu_1=\pi/2$, $\mu_2=2.40483$, $\mu_3=\pi$)
and the constants are $a_d = 1, 3.71038, \pi^2$ in $d = 1, 2, 3$,
respectively~\cite{Luck89}.
Substituting this approximative solution for the  saddle point into
the action~(\ref{eq:Snewresc}) leads immediately to the DOS tail of
the form~\cite{Shapiro12,Giacomelli10,Luck89}
\begin{eqnarray}\label{eq:dens-saddle}
\nu\left(E\right)= {\cal A}_d~ \exp{\left[-
v_d~\left({E_\xi}/{E}\right)^{\frac{d}{2}}~\ln{\left({I_0}/{E}\right)}
\right]},
\end{eqnarray}
with $v_d=\mu_d^d~\pi^{d/2}/\Gamma(d/2+1)$.
The derivation of the saddle point solution can be simplified
in one dimension. The basic idea is to treat Eq.~(\ref{eq:Snewresc}) as
the classical action
\begin{eqnarray}
S_{\rm cl}=  \int d {x} \left\{ \left[\nabla_{{x}}
{\phi_0}({x})\right]^2
-{\cal U}(\phi_0)
\right\},
\label{eq:Snewrescbis}
\end{eqnarray}
of a particle moving in the potential
${\cal U}(\phi_0)= {\tilde{E}} {\phi_0}^2({x})
- \ln \left[1+ s {{\phi_0}}^{2}(x)\right]$
with space coordinate $\phi_0$ and time $x$.
Since the system is conservative the energy of the particle
\begin{eqnarray}\label{eq:firstordernewresc}
{\cal E}_0= \dot{\phi_0}^2 + {\cal U}(\phi_0)
\end{eqnarray}
is constant along any trajectory.
The saddle point solution corresponds to the particle trajectory at
${\cal E}_0=0$.
By using a simple variable transformation the action~(\ref{eq:Snewrescbis}) can
be rewritten  as
\begin{eqnarray}\label{eq:resratiooned}
{\cal S}_{\mathrm{cl}} =2 \int\limits_0^{z_0} d z
\sqrt{z^{-1}\ln (1+s ~z)-\tilde{E}},
\end{eqnarray}
where $z_0$ corresponds to zero of the expression under the root.
For $\tilde{E}\rightarrow 0$ the asymptotic behavior of Eq.~(\ref{eq:resratiooned})
is ${\cal S}_{\mathrm{cl}} \approx \pi \sqrt{{1}/\tilde{E}}%
\ln{\left({s}/\tilde{E}\right)}$
in agreement with the DOS~(\ref{eq:dens-saddle}) in one dimension.

\section{Fluctuations around the saddle point}
\label{sec:fluct}

In this section we extend the instanton approach in order
to calculate the dependence of the prefactor ${\cal A}_d$ upon $E$.
To this end we expand the action~(\ref{eq:action-av})
around the saddle point $\bar\phi=\bar\phi_{\mathrm{cl}}+\bar\phi'$
to second order in the fluctuation fields.
The instanton contribution to the Green function
is then given by
\begin{eqnarray}
\mathcal{G} (x,x';E)
&\sim & \int \mathcal{D} \phi'\ \phi_{\mathrm{cl}1}(x) \phi_{\mathrm{cl}1}(x')\nn \\
&& \ \ \ \ \times \exp[-\frac12 \int d^d x {\phi}_{\alpha}'
M_{\alpha \beta}{\phi}_{\beta}' ].\label{eq:pref}
\end{eqnarray}
Assuming the saddle point solution of the form~(\ref{eq:phi-clas})
the operator $M_{\alpha \beta}$ can be diagonalized  using
the longitudinal and transverse projector operators in the replica space as
\begin{eqnarray}\label{eq:M}
{M}_{\alpha \beta} = M_\mathrm{L} n_{\alpha} n_{\beta}
+M_\mathrm{T} \left(\delta_{\alpha \beta}-n_{\alpha} n_{\beta}\right).
\end{eqnarray}
The transverse and longitudinal operators can be written in the form
\begin{eqnarray} \label{eq:MT-ML}
M_{T,L}=-\nabla^2 + U_{T,L} +m^2
\end{eqnarray}
where we have defined the mass $m^2=s-\tilde{E}$
such that the potentials
\begin{eqnarray}
U_{T}(r)&=& \frac{s}{1+s ~{\phi_0}^2} - s ,\label{eq-potentials-UT-1} \\
U_{L}(r)&=& \frac{s}{1+s ~{\phi_0}^2} - s -
\frac{2 s^2~ {\phi_0}^2}{\left(1+s~ {\phi_0}^2 \right)^2} \label{eq-potentials-UL-1}
\end{eqnarray}
vanish at infinity.
Note that the  transverse projector operator
has $(N - 1)$ zero modes corresponding to invariance under $O(N)$
rotations in the replica space while  the longitudinal operator
has  $d$ zero modes corresponding to translational invariance.
In order to obtain a finite result from the Gaussian integration
in Eq.~(\ref{eq:pref}), the zero modes of the
operators $M_T$ and $M_L$ have to be separated and integrated out exactly
without using the Gaussian approximation.
To that end one can perform
transformation to a collective coordinates $x_0$ and
$\bar{n}$~\cite{Cardy78,Langer67}.
This yields
\begin{eqnarray}\label{eq:densgen}
\mathcal{G} (x,x';E)&
 \sim & \int d^d x_0~d \bar{n}~{{ J}^{t}}~ \phi_{0}(x-x_0) \phi_{0}(x'-x_0) \nn\\
&& \times\int \mathcal{D}\bar{\phi}' \exp[-\frac12 \int d^dx {\phi}_{\alpha}'
M^{\prime}_{\alpha \beta}{\phi}_{\beta}'], \label{eq-pref-2}
\end{eqnarray}
where ${J}^{t}$ is the Jacobian of the transformation to the collective coordinates
$x_0$ and $\bar{n}$ and the prime in $M^{\prime}_{\alpha \beta}$ means
that the zero modes have been omitted.
The Jacobian calculated to leading order in the energy  $\tilde{E}$
by expanding the model in the fields around the minimum
is given by~\cite{Christ75}
\begin{eqnarray}
J^t\sim\left[
\int d^d x \left(\nabla \bar\phi_{0}\right)^2
\right]^{d/2}
\left[
\int d^d x ~{\bar\phi_{0}}^2
\right]^{(N-1)/2}. \label{eq-Jacob1}
\end{eqnarray}
The functional integral in Eq.~(\ref{eq-pref-2}) contributes with
\begin{eqnarray}\label{eq:ratio}
 \int \mathcal{D} \bar{\phi}' &&\exp{\left[-\frac12 \int d x {\phi}_{\alpha}'
{M}^{\prime}_{\alpha \beta}{\phi}_{\beta}'\right]} \nn \\
&&=\det{'\tilde{M}_L}^{-1/2} \cdot \det{'\tilde{M}_T}^{-(N-1)/2}
\end{eqnarray}
where $\det{'}$ stands for the product of all non-zero eigenvalues including
the continuous part of the spectrum.
The operators $M_T$ and $M_L$ given by Eq.~(\ref{eq:MT-ML})
are Schr{\"o}dinger-like operators. Their
spectrum turns out to be very sensitive to the precise form of the
saddle point solution of
Eq.~(\ref{eq:schroedingerphinew}), which is not known analytically even
in one dimension.
Unfortunately, the dependence of the eigenvalues on the energy parameter $\tilde{E}$
cannot be easily extracted using simple scaling arguments as in the case of
Gaussian uncorrelated
disorder (see Appendix~\ref{appendix-C}). Nevertheless, there are methods which
allow one to calculate the functional determinants with excluded zero modes  even
without knowing precisely the spectrum.

\subsection{One dimensional case}

\begin{figure}
\begin{center}
\includegraphics[width=7.5cm]{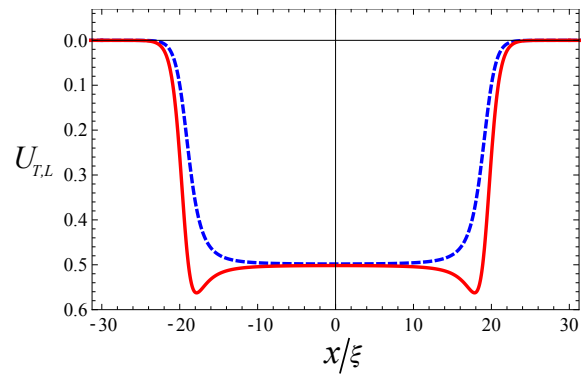}
\end{center}
 \caption{One-dimensional potential wells $U_T$ (blue dashed)
 and $U_L$ (red solid)  for $s=0.5$ and  $E=0.01$. }
 \label{fig:potential}
\end{figure}

We start with the one dimensional case.
The potentials $U_T$ and $U_L$ corresponding to
the operators  $M_T$ and $M_L$
obtained from numerical solution of the saddle point equation in $d=1$
are shown in Fig.~\ref{fig:potential}.
In the low energy limit $E\rightarrow 0$, we find that
the potentials $U_T$ and $U_L$ approach asymptotically
a square potential well of width  ${\pi}\sqrt{1/{\tilde E}}$ and
depth $s\mp {\tilde E}\left(1/\ln{ \frac{s}{\tilde E}}\right)$ respectively.
The operator $M_T$ has one zero mode corresponding to
the lowest symmetric state.
The operator $M_L$ has the only zero energy state corresponding to
the lowest antisymmetric state while its lowest symmetric state
has a negative eigenvalue which gives a non-zero contribution to
the imaginary part of the Green function.

When the spectrum of the Schr{\"o}edinger operator is known analytically the
zero mode can be explicitly excluded from the product of eigenvalues.
Therefore,
the determinant can be calculated
simply as an infinite product of non-zero eigenvalues. This is illustrated
in Appendix~\ref{appendix-C1} for the fluctuation operators $M_T$ and $M_L$
arising in the Gaussian disorder model~\cite{Cardy78}.
For the blue detuned speckle
the determinants of $M_T$ and $M_L$ cannot be calculated
simply as a product of non-zero eigenvalues because
the spectrum cannot be found analytically.
Fortunately, Gel'fand and
Yaglom (GY)~\cite{Gelfand60,Kleinert04} derived long ago a general formula which
allows one to calculate the functional determinant
of a Schr{\"o}dinger like operator at least in one dimension without knowing
any of its eigenvalues.
The GY method can be applied to an operator of the form
\begin{eqnarray}\label{eq:Sch-operator}
{\cal M}=- \frac{d^2 }{dx^2} + U(x) + m^2,
\end{eqnarray}
which is defined on $x\in[-L,L]$ for the wave functions satisfying
the boundary conditions $u(-L)=u(L)=0$.  The limit $L \to \infty$ can
be taken at the end of the calculation.
Since the well defined object is rather a ratio of two determinants than a single
functional determinant itself
it is convenient to
introduce a free operator ${\cal M}_{\rm free}=- \frac{d^2 }{dx^2}  + m^2$.
The GY theorem~\cite{Gelfand60} states that
\begin{eqnarray}\label{eq:GY}
\frac{\det{\cal M}}{\det{\cal M}_{\rm free}}=\frac{u(L)}{u_{\rm free}(L)},
\end{eqnarray}
where $u(x)$ and $u_{\rm free}(x) $ are the solutions of the Cauchy problems
\begin{eqnarray}\label{eq:eigen_probl-1}
{\cal M} u(x)=0 \hspace{5mm} \mathrm{and } \hspace{5mm} {\cal M}_{\rm free}
u_{\rm free}(x)=0
\end{eqnarray}
with the initial conditions:
\begin{eqnarray}\label{eq:eigen_probl}
 u(-L)=u_{\rm free}(-L)=0, \ \ \ u'(-L)=u_{\rm free}^{\prime}(-L)=1. \nn \\
\end{eqnarray}
Due to the presence of eigenfunctions with zero eigenvalue, whose contributions to
the determinant have to be excluded, the
GY formula~(\ref{eq:GY}) has to be slightly modified for the operators
$M_T$ and $M_L$.
A simple regularization consists of introducing an infinitesimal shift of the
spectrum by a small shift of the mass $m$.
Then, the original determinant  with the excluded zero mode
can be derived by differentiating with respect to the mass.
This method is illustrated for the operators $M_T$ and $M_L$
of the Gaussian disorder model in Appendix~\ref{appendix-C2}.

In the case of the speckle potential it turns out to be more convenient
to use another regularization approach
which has been recently proposed by Tarlie and McKane (TM) in Ref.~\cite{Tarlie95}.
It is based on the GY method generalized to an arbitrary boundary
conditions by Forman in Ref.~\cite{Forman87}. The basic idea is to
regularize the determinant by modifying  the boundary
conditions. This changes the zero eigenvalue to a nonzero one which can be estimated
to lowest order in the difference between the original and regularized
boundary conditions.
Assuming that the zero mode of the operator $\mathcal{M}$ is given by
$v_0(x)$ the ratio of the two determinants with excluded zero mode
can be written as
\begin{equation}\label{eq:limitdet-0}
\frac{\det' \mathcal{M}}{\det \mathcal{M}_{\rm free}}=-
\frac{\left\langle v_0 |v_0 \right\rangle}{v_0'(-L) v_0'(L)  }
\frac{u^{\prime}_{\rm free} (-L)}{u_{\rm free} (L)},
\end{equation}
where we defined the scalar product
\begin{eqnarray}
 \left\langle v_0 |v_0 \right\rangle := \int_{-L}^{L} dx \, v_0^2(x).
\end{eqnarray}
The TM formula is very useful because the zero mode of the operator
$M_T$ is given by  the classical solution
$\phi_0(x)$ while for the operator $M_L$ the zero mode
is simply given by its derivative $\phi_0^{\prime}(x)$.
This is true not only in the case of the one dimensional speckle potential but also
for the problem with uncorrelated Gaussian disorder where
the TM formula is shown to reproduce the correct result of Cardy
(see Appendix~\ref{appendix-C3}).
By inserting the zero mode solutions
in Eq.~(\ref{eq:limitdet-0}) the ratio of the two determinants $M_T$ and $M_L$
of Eq.~(\ref{eq:ratio}) in one dimension can be rewritten as
\begin{equation}\label{eq:limitdet}
\frac{\det'M_T}{\det' M_L}=\lim_{L\rightarrow\infty}
\frac{\left\langle\phi_{0}|\phi_{0}\right\rangle}{\left\langle
\phi^{\prime}_{0}|\phi^{\prime}_{0}\right\rangle}
\frac{\phi^{\prime\prime}_{0}(-L)
\phi^{\prime\prime}_{0}(L)}{\phi^{\prime}_{0}(-L) \phi^{\prime}_{0}(L)},
\end{equation}
because the contribution from the free operator cancels out.
Moreover, in one dimension,
the derivatives of the classical solution can be easily obtained from the first
order differential
equation~(\ref{eq:firstordernewresc}) derived from the analogy with the
particle in a conservative potential.
Although the function $\phi_0(x)$ is known only numerically,
the limit of Eq.~(\ref{eq:limitdet}) can be calculated analytically by using that
this solution is regular at infinity.
We find the exact relation
\begin{equation}\label{eq:detspeckle}
\frac{\det'M_T}{\det' M_L}=%\lim_{L\rightarrow\infty}
- \frac{\left\langle\phi_{0}|\phi_{0}\right\rangle}{\left\langle
\phi^{\prime}_{0}|\phi^{\prime}_{0}\right\rangle}
\left(s-\tilde{E}\right).
\end{equation}
When we substitute this result in the DOS~(\ref{eq:densgen}) the ratio of two scalar
products in Eq.~(\ref{eq:detspeckle})
cancel exactly the Jacobian~(\ref{eq-Jacob1}).
Using the saddle point solution~(\ref{eq:newresc-f}) in the asymptotic
limit $E\rightarrow 0$ we obtain
\begin{eqnarray}\label{eq:densspeckleB}
\mathcal{A}_1
 \sim \frac{\ln{\left({s}/\tilde{E}\right)}}{\tilde{E}^{3/2}}{s}^{1/2}.
\end{eqnarray}
By collecting all contributions, we find the tail of the DOS in one dimension
\begin{eqnarray} \label{eq-fit}
\nu = \frac{A}{\xi I_0}  \left(\frac{I_0}{E}\right)^{3/2} \ln
\left( \frac{I_0}{E}\right)
\exp\left[ - \pi s^{-1/2} \sqrt{\frac{I_0}{E}} f_1\left(
\frac{I_0}{E}\right) \right], \nn \\
\end{eqnarray}
where $A$ is a numerical constant and
\begin{equation}
f_1(y) = \frac{2}{\pi} \int\limits_0^{z_0} d z
\sqrt{z^{-1}\ln (1+y z)-1},
\end{equation}
whose asymptotics for large $y$ is  $f_1(y)=\ln y$.
\begin{figure}
\begin{center}
\includegraphics[width=7.5cm]{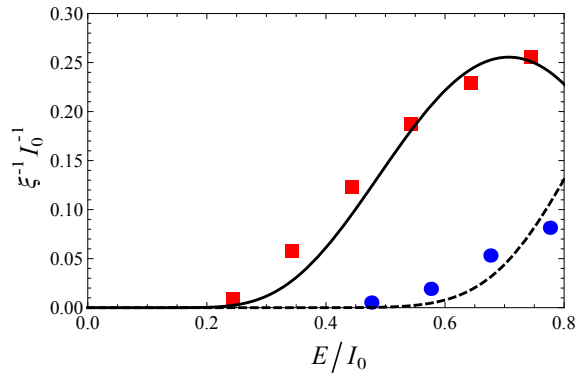}
\end{center}
 \caption{The one particle DOS for a blue-detuned
speckle potential computed numerically for $s = 1$ (red squares)
and $s=0.1$ (blue circles). The solid and dashed lines are the best
fit of the tails by Eq.~(\ref{eq-fit}).  }
\label{fig:fitting}
\end{figure}
In order to check this formula we have fitted the low energy tail  of the DOS
in 1D speckle potential
which we computed numerically using the exact Hamiltonian diagonalization in
Ref.~\cite{Giacomelli10}. The result is shown in Fig.~\ref{fig:fitting}.

\subsection{Higher dimensions}
\label{sec:GY-d}

The GY method can be extended to determinants of the
Schr\"{o}dinger-like
operators in $d$ dimensions in the case of radially symmetric
potentials~\cite{Kirsten06}.
Due to the radial symmetry of the operators, their eigenfunctions can
be decomposed into a product of radial parts and  hyperspherical harmonics
\begin{eqnarray}
\Psi\left(r,{\mathbf\vartheta}\right)=\frac{1}{r^{\frac{d-1}{2}}}
\psi_{\ell}(r)~Y_{\ell}(\vartheta).
\end{eqnarray}
The radial parts $\psi_{\ell}(r)$ are then solutions to the radial equations
\begin{eqnarray} \label{eq-rad-operators-1}
 {M}_{T,L}^{(\ell)}&&~\psi_{\ell}(r):=
\left( -\frac{d^{2}}{dr^{2}}+
\frac{\left(\ell+\frac{d-3}{2}\right)\left(\ell+\frac{d-1}{2}\right)}{r^{2}}
+\right.\nonumber\\
&&\left. +U_{T,L}(r) + m^2
\right){\psi_{\ell}(r)}=\lambda~\psi_{\ell}(r). \
\end{eqnarray}
The radial eigenfunctions $\psi_{\ell}(r)$ come with a degeneracy
factor of
\begin{eqnarray} \label{eq-deg-1-0}
{\rm deg}\left(\ell;d \right)=\frac{(2\ell +d-2)(\ell+d-3)!}{\ell !(d-2)!}.
\end{eqnarray}
The determinant of a radially separable operator
can be calculated by combining the determinants for each partial wave $\ell$
with the weights given by the degeneracy factor~(\ref{eq-deg-1-0}) as follows
\begin{eqnarray}\label{eq:comb-l}
\ln{\frac{\det'{M}_{T,L}}{\det{M}_{\rm free}}}=
\sum_{\ell=0}^{\infty}{\rm deg}\left(\ell;d \right)
~\ln{\frac{\det'{M}_{T,L}^{(\ell)}}{\det{M}_{\rm free}^{(\ell)}}},
\end{eqnarray}
where the free operators have been defined as
\begin{eqnarray}
{M}_{\rm free}^{(\ell)}=-\frac{d^{2}}{dr^{2}}+
\frac{\left(\ell+\frac{d-3}{2}\right)\left(\ell+\frac{d-1}{2}\right)}{r^{2}}
+m^2.
\end{eqnarray}
In order to compute the determinants of the partial operators in Eq.~(\ref{eq:comb-l})
one can use the GY method for those determinants that have no zero modes
and the MT method for those that have them.  The only partial operators which have
zero modes are ${M}_{T}^{(0)}$ and ${M}_{L}^{(1)}$.
All other determinants can be computed using the GY formula
\begin{eqnarray}\label{eq:G-Y3d}
\frac{\det{M}_{T,L}^{(\ell)}}{\det{M}_{\rm free}^{(\ell)}}
=\lim_{R\rightarrow \infty}\frac{u^{(\ell)}(R)}{u_{\rm free}^{ (\ell)}(R)},
\end{eqnarray}
where $u^{(\ell)}$ and $u_{\rm free}^{(\ell)}$
are the solution of the following Cauchy problems
\begin{eqnarray}
&& {M}^{(\ell)}u^{(\ell)}(r)=0 , \hspace{3mm} u^{(\ell)}
  \sim r^{\ell+\frac{(d-1)}{2}}, \hspace{3mm} r\rightarrow 0,\label{eq:bound-cond-1} \\
&& {M}^{(\ell)}_{\rm free}u^{(\ell)}_{\rm free}(r)=0 , \hspace{3mm}
  u_{\rm free}^{(\ell)}\sim r^{\ell+\frac{(d-1)}{2}}, \hspace{3mm}
r\rightarrow 0. \label{eq:bound-cond-2} \ \ \ \ \
\end{eqnarray}
The determinants of the operators ${M}_{T}^{(0)}$  and  ${M}_{L}^{(1)}$
with excluded zero modes are given by the generalization of
Eq.~(\ref{eq:limitdet-0}) which reads
\begin{eqnarray}
\label{eq:limitdet-3}
\frac{\det' \mathcal{M}}{\det \mathcal{M}_{\rm free}}=-
\lim\limits_{R\to \infty}
\frac{\left\langle v_0 |v_0 \right\rangle}{ v_0'(R) u_{\rm free} (R)  }.
\end{eqnarray}
Here we have used that the zero modes $v_{0}(r) $ of the operators
${M}_{T}^{(0)}$ and ${M}_{L}^{(1)}$ and the solution $u_{\rm free}(r)$
of the Cauchy problem (\ref{eq:bound-cond-2})
have the same behavior at $r\to 0$.
\begin{figure}
\begin{center}
\includegraphics[width=83mm]{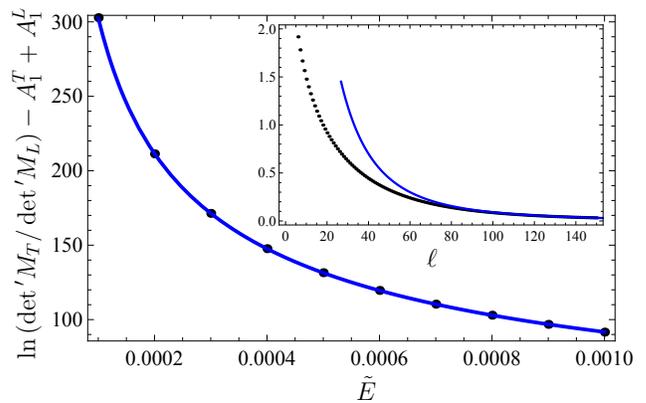}
\end{center}
 \caption{The logarithm of the ratio of determinants $M_T$ and $M_L$ with
 subtracted diagrams $A^{T,L}_1$ computed numerically for $s=0.5$ and $d=2$
 as a function of $\tilde{E}$ using Eq.~(\ref{eq:sum-finite}).
 The solid blue line is a fit by
 ${3.18156}/\sqrt{\tilde{E}}+2.82111 \ln \tilde{E}+10.5708$.
(Inset) The logarithm of the ratio of partial determinants for
$d=2$, $\tilde{E}=0.0005$, and $s=0.5$ as a function of $\ell$. The solid blue
line is the asymptotic behavior ${162085}/{(\ell+21.4092)^3}$.
 }
\label{fig:determinant}
\end{figure}
It turns out that the sum over $\ell$ in Eq.~(\ref{eq:comb-l}) diverges
for $d\ge2$ as $1/\ell$. This divergence is a general property of functional
determinants in $d\ge 2$~\cite{Dunne08} that was not discussed
in the most works
on the instanton approach to the DOS of disordered
systems~\cite{Cardy78,John84,Luttinger83,yaida12}.
This divergence reflects the fact that the field
theory~(\ref{eq:Snewresc})
has to be renormalized~\cite{bresin1980}
that is beyond the scope of the present paper.
To make the theory finite we introduce the UV cutoff $\Lambda$ and
separate the divergencies from the sum over $\ell$ in Eq.~(\ref{eq:comb-l}).
Following Ref.~\cite{Kirsten06} we use the diagrammatic representation
of the determinants ${M}_{T,L}$  explained in Appendix~\ref{sec:app:1} and
given by Eq.~(\ref{eq16-1})
with $V(r) = U_{T,L}(r)$. For $2 \le d < 4$ the only divergent diagram is $A_1$
which is linear in $V$. For $d=4$ one has also subtract the diagram $A_2$ which
is quadratic in $V$.
Subtracting the divergent diagram from the partial determinants we obtain
\begin{eqnarray}
 \ln \frac{\det' {M}_{T,L} }{
  \det {M}_{\rm free}} &=& A_1 + \sum_{\ell=0}^\infty {\rm deg}(\ell; d)\,
    \left[ \ln \frac{\det' {M}^{(\ell)}_{T,L} }{
  \det {M}^{(\ell)}_{\rm free}} \right]_{\emptyset(V)}, \nn \\
  \label{eq:sum-finite}
\end{eqnarray}
where the symbol $\left[...\right]_{\emptyset(V)}$ means that the parts of order $V$
have been subtracted.
The sum over $\ell$ in Eq.~(\ref{eq:sum-finite})
becomes finite because the UV divergences have been accumulated in the
regularized Feynman diagram $A_1$. The explicit form of the terms
which have to be subtracted for a general potential $V(x)$ is
given by the expansion~\cite{Kirsten06}
\begin{eqnarray}
 \ln \frac{\det \mathcal{M}^{(\ell)} }{
  \det \mathcal{M}^{(\ell)}_{\rm free}} &=& \int\limits_0^{\infty}dr\, r\, V(r)
K_{\nu} (m r) I_{\nu} (m r) \nn \\
&&-\int\limits_0^{\infty}dr\,r\, V(r)
K_{\nu}^2 (m r) \int\limits_0^rdr'\, r' \, \nn\\
&& \times V(r') I_{\nu}^2 (mr')+ {\cal O} (V^3) \quad ,
\label{jost-expansion}
\end{eqnarray}
where $I_{\nu}$ and $K_{\nu}$ are the Bessel function with
$\nu\equiv l+\frac{d}{2}-1$.
Thus, one can compute the regularized ratio of the functional determinants
(\ref{eq:sum-finite})  by solving numerically the Cauchy
problems~(\ref{eq:bound-cond-1})-(\ref{eq:bound-cond-2}) and
using the GY and MT formulas (\ref{eq:G-Y3d}) and (\ref{eq:limitdet-3}).
We have found that after subtracting the diverging term given by the first
line in Eq.~(\ref{jost-expansion})
the sum over $\ell$ in Eq.~(\ref{eq:sum-finite})  converges
asymptotically as $1/\ell^3$ in
$d=2$ and as $1/\ell^2$ in $d=3$. As an example, this sum is shown as a
function of $\tilde{E}$ for $d=2$ and $s=0.5$  in Fig.~\ref{fig:determinant}.
The total ratio of the determinants, however, is dominated by the exponential of
\begin{eqnarray}
A_1^{T}-A_1^{L}=\int_q \frac1{q^2+m^2} \int d^d x [U_T(x)-U_L(x)].
\end{eqnarray}
Neglecting a power-law correction resulting from the Jacobian~(\ref{eq-Jacob1})
we arrive at
\begin{eqnarray} \label{eq-cor2}
\mathcal{A}_d(E)\sim
\exp\left[\gamma_d\left (\frac{I_0/E}{\ln(I_0/{E})}
\right)^{(d-1)/2}\right]
\end{eqnarray}
with UV cutoff-dependent coefficients $\gamma_2=1.89\ln({\Lambda}/{s})$ and
$\gamma_3=14.81 \Lambda$.

\section{Weakly interacting Bose gas in speckle potential}
\label{sec:interactions}

We now consider the effect of weak repulsive interaction on the bosons
in the Lifshitz tail. We restrict our consideration  to the three dimensional
system in dilute regime. The corresponding Hamiltonian has the well-known
form~\cite{Falco09-2}
\begin{eqnarray}
H &=& \int d^3 x  \psi^{\dagger}(x) \left(-\frac{\hbar^2}{2m_0}\nabla^2
+ V(x)-\mu \right )\psi(x)  \nn \\
 && \ \ \ \ \ + \, \frac{g}{2} \int d^d x (\psi^{\dagger}(x) \psi(x))^2,
\end{eqnarray}
where $\psi(x)$ is the secondary quantized wave function, $V(x)$ - random
potential and $\mu$ - the chemical potential.
The positive coupling constant is given by $g=4\pi \hbar^2 a_s/m_0$,  where $a$
is the scattering length and we assume a low concentration
of bosons $n$,
such that $n a_s^3 \ll 1$. The aim is to find how the bosons  fill the
random potential when we add the particles one by one. In particular,
we are interested in the dependence of the chemical potential on the density of
bosons. It is instructive to compare the case of a bounded from below potential
with the random uncorrelated Gaussian potential studied by one of us
in Ref.~\cite{Falco09-2}.

In the case of Gaussian disorder  the asymptotic behavior of the
DOS for large negative energy $E$ is dominated by the optimal wells of width $R$
with the energy  $E_0(R) = - \hbar^2/(2m_0 R^2)$ which decreases with shrinking of
$R$.
The density of optimal
wells, which can be found from the corresponding instanton solution, is
$n_w(R) \sim e^{-\mathcal{L}/R}/R^3$, where $\mathcal{L}=\hbar^4/(m_0^2 \gamma^2)$
is the so-called Larkin length related to the strength of
disorder $\gamma$~\cite{Falco09-2}.
In the presence of weak repulsive interactions
the positive repulsion energy per particle grows
with decreasing $R$ as $E_r (R) = 3 g \mathcal{N}(R)/(4\pi R^3)$, where
$\mathcal{N}(R)=n/n_w(R)$ is
the typical number of bosons in optimal wells.
Since the both energies  $E_0(R) $  and $E_r(R)$
have opposite behavior with respect to decreasing $R$  one has to optimize
the total energy in order to find the size of the optimal well renormalized
by interactions. This yields with the
logarithmic precision $R(n)= \mathcal{L}/\ln(n_c/n)$.
Then the relation between the chemical potential and the density is given
by~\cite{Falco09-2}
\begin{eqnarray}
\mu (n) = -\frac{\hbar^2}{2 m_0 R^2(n)} \approx -
\frac{\mathcal{E}}2 \left(\ln \frac{n_c}{n}\right)^2,
\end{eqnarray}
where $\mathcal{E}=\hbar^2/(m_0 \mathcal{L}^2)$ and
$n_c=(3 \mathcal{L}^2 a_s )^{-1}$.

In the case of the speckle potential both the energy corresponding to the optimal
well $E_0(R)=\hbar^2/(2m_0 R^2)$  and the positive repulsion energy
decay with growing the size of the typical well. Thus, there is no competition
between the disorder and interactions so that we have no need for optimization
of the total energy.
Neglecting the kinetic energy
and using Eq.~(\ref{eq:dens-saddle}) to estimate the
density of the optimal wells we obtain with the logarithmic precision
\begin{eqnarray} \label{eq-mu-1}
\mu (n) \approx E_{\xi}  \left( v_3 \ln\frac{I_0}{E_\xi}\right) ^{2/3} \,
\left(\ln \frac{n_0}{n}\right)^{-2/3},
\end{eqnarray}
where $n_0=(6 \xi^2 a_s )^{-1}$ and $v_3=4\pi^4/3$.
The asymptotic behavior~(\ref{eq-mu-1}) holds
for  $n\ll n_0$ and $E_\xi \ll I_0$ and is an agreement
with Ref.~\cite{Gopalakrishnan12}.

\section{Conclusion}

We have studied the low energy behavior of the DOS for
non-interacting bosons in a $d$ dimensional blue detuned laser speckle
potential.   We have shown  that for $E \le E_\xi$
the precise form of the electric field correlator does not affect the
asymptotic  behavior. Using an instanton approach we have found the saddle point
solution which gives the leading exponential behavior.
Integrating out the Gaussian fluctuation around this solution we have
expressed the prefactor in the form of ratio of two functional determinants.
In one dimension we calculated the ratio of functional determinants exactly
using the generalized GY method which allows one to take into account not only
the discrete part of the spectrum of fluctuation operators but
also the continuous one. In higher dimensions the corresponding ratio diverges
that has been overlooked in most of the previous work on the instanton approach
to the DOS of disordered systems. Using the partial wave decomposition we can
separate the UV divergences to a regularized one-loop Feynman diagram and
obtain a finite result for the DOS tail in $d>1$. In
Appendix~\ref{sec-GY-higher-dimensions} we show that this method gives
a correct result for the case of Gaussian uncorrelated disorder.
We also discussed the effect of weak repulsion interactions in the DOS tail.
In contrast to the Gaussian unbounded disorder the interactions and disorder do
not compete and the optimal wells are not renormalized by
interactions that leads to a different dependence of the chemical potential on
the bosons density.

\begin{acknowledgments}
We would like to thank Thomas Nattermann, Valery Pokrovsky and Boris Shklovskii
for useful discussions. AAF acknowledges
support by ANR grants 13-JS04-0005-01 (ArtiQ) and 2010-BLANC-041902 (IsoTop).
\end{acknowledgments}

\appendix

\section{Functional determinant representation}
\label{sec:app:1}

The diagrams in Eq.~(\ref{eq16-1})
can not be summed  up  for an arbitrary distribution of the electric fields
including those that appear in real experiments.
In this appendix we show that
the sum can be performed
for a special choice of the Gaussian distribution of the electric fields
with zero mean and variance
\begin{eqnarray}
G(x)=I_0\frac{2^{d/2} }{\Gamma
   \left(1-\frac{d}{2}\right)}\left(\frac{x}{\xi }\right)^{1-\frac{d}{2}}
   K_{d/2-1}\left(\frac{x}{\xi }\right),\label{eq16-0}
\end{eqnarray}
where $K_{\nu}(x)$ is the  modified  Bessel function. A similar
correlator appears in the problem of the Bragg glass
studied in Ref.~\cite{Fedorenko14}.
For $d=1$ the variance~(\ref{eq16-0})
reduces to $G(x)= I_0  \ e^{-|x|/\xi}$.
This is particular interesting because the asymptotic behavior of the DOS does not
depend on the precise form of correlations in disorder but it is determined by the
lower energy states which spread over distances larger than the disorder correlation
length $\xi$.
Therefore, in order to study the lowest order corrections to the asymptotic tail
due to presence of correlations one can use Eq.~(\ref{eq16-0}) as a reasonable
approximation for the variance of the electric field.

The starting point is the following diagrammatic representation of the
ratio of two functional determinants
\begin{eqnarray}
&&\ln\left(\frac{\det ( - \nabla^2 + V(x) + m^2 ) }{
  \det ( - \nabla^2 + m^2 )}\right)= \sum\limits_{n=1}^{\infty}   \frac{(-1)^{n+1}}{n}A_{n}  \nn \\
&& \mbox{}\hspace{10mm}
=
\parbox{8mm}{\includegraphics[width=8mm]{diag1}} -
\frac{1}{2}
\parbox{8mm}{\includegraphics[width=8mm]{diag2}} +
\frac{1}{3}
\parbox{8mm}{\includegraphics[width=8mm]{diag3}} - ...
\label{eq16-1}
\end{eqnarray}
In the diagrams shown in Eq.~(\ref{eq16-1}) the dots correspond to the
potential $V(x)$
and the lines stand for the Green's function
\begin{eqnarray}\label{eq-G-1}
C_0(x)=\frac{m^{d-2}}{(2\pi)^{d/2}}\frac{K_{d/2-1}(m |x|)}{(m |x|)^{d/2-1}},
\end{eqnarray}
which satisfies the equation
\begin{eqnarray}
\left [-\nabla^2 + m^2\right] C_0(x) = \delta(x) \label{eq:Greens-1}.
\end{eqnarray}
If we separate the combinatorial factors from the diagrams in
Eq.~(\ref{expEE}) we obtain the same series as in Eq.~(\ref{eq16-1}).
Thus, one can formally rewrite the sum of the diagrams in Eq.~(\ref{expEE})
as a ratio of two functional determinants
with the mass $m=1/\xi$ and the potential $V(x)= \frac12 I_0 \bar{\phi}^{2}(x)/C_0(0)$
resulting from a random Gaussian electric field with zero mean and variance
$G(x)=I_0 C_0(x)/C_0(0)$.
Note that divergency of $C_0(0)$ when $d\ge 2$ reflects the fact that the functional
determinants in Eq.~(\ref{eq16-1}) require a renormalization for $d\ge2$. Here
we restrict ourselves to the case $d=1$ and obtain
\begin{eqnarray}
&& \!\!\!\!\!\!
 S(E)= \int d x \left[  \frac{\hbar^2}{4m_0^2}(\nabla \bar{\phi}(x))^2 - \frac12 E \bar{\phi}^{2}(x) \right] \nonumber \\
&& \ \ \ +\ln\left(\frac{\det ( - \nabla^2 + \left(I_0/\xi\right)
\bar{\phi}^{2}(x) +  1/\xi^2 ) }{
  \det ( - \nabla^2 + 1/\xi^2 )}\right). \label{eq:16-3}
\end{eqnarray}
In the limit $\xi\to 0$ we can neglect the $\nabla^2$-operator in the
determinants of Eq.~(\ref{eq:16-3}) and recover Eq.~(\ref{eq:action-av})
using that $\ln \det = \mathrm{Tr} \ln$.
The corresponding saddle point equation
\begin{eqnarray}
 \frac{\hbar^2}{2m_0^2} \nabla^2 \bar{\phi}(x) &+& E \bar{\phi}(x) = \frac{\delta}{\delta \bar{\phi }(x)}
\ln \det \left[ - \nabla^2  \right.\nn \\
&& \left.+ \left(I_0/\xi\right)
\bar{\phi}^{2}(x) +  1/\xi^2 \right]
\end{eqnarray}
has the form of a gap equation known in relativistic
quantum field theory~\cite{Dunne08}.

\section{Variational method with Gaussian trial functions }
\label{sec:app:2}

One can also sum  up all diagrams in Eq.~(\ref{expEE}) for
a particular class of functions  $\bar{\phi}(x)$ that can be used to
find variationally
an approximative instanton solution by means of the trial function method.
Let us assume that the one dimensional electric field correlator has the form
\begin{eqnarray}
G(x-y)=I_0 \xi \delta_{\xi}(x-y) \hspace{2mm} \mathrm{with } \hspace{2mm}
\delta_{\xi}(x) = \frac1{\xi } e^{-\pi x^2/\xi^2}.
\end{eqnarray}
For the trial function we consider $\phi(x)= \bar{n} \phi_0(x)$ with $\bar{n}^2=1$ and
\begin{eqnarray} \label{eq-trial-fun}
\phi_0(x)= \sqrt{C} e^{-\pi x^2/2 a^2}.
\end{eqnarray}
By substituting the trial function into the potential part of the
action~(\ref{expEE}) we find that the diagram with $n$ electric field correlators
reads
\begin{eqnarray}
\parbox{17mm}{\includegraphics[width=16mm]{diagN}} &=&
 (n-1)!\, I_0^n\,  C^n
\int dx_1...dx_n  \exp\left[-\frac{\pi x_1^2}{a^2}
 \right. \nonumber \\
  && -\frac{\pi (x_1-x_2)^2}{\xi^2} ...
 - \frac{\pi (x_{n-1}-x_n)^2}{\xi^2} \nn \\
&&
\left.-\frac{\pi x_n^2}{a^2}
-\frac{\pi (x_n-x_1)^2}{\xi^2} \right]. \label{eq-diag-20}
\end{eqnarray}
Upon making the variable rescaling  $x_i \to \xi x_i/\pi $,  the integral in
Eq.~(\ref{eq-diag-20}) can be rewritten as
\begin{eqnarray}
 \frac{\xi^n}{\pi^n}
\int dx_1...dx_n
 \exp\left[-\sum\limits_{i,j=1}^n x_i A_{ij} x_j \right]
 = \frac{\xi^n}{\pi^{n/2}}  [\det A_n]^{-1/2}, \nn \\
\end{eqnarray}
where we have introduced the matrix
\begin{eqnarray}
A_n=\left(
  \begin{array}{cccccc}
    2+\epsilon^2   &  -1          &  0           &  ...       &  0         & -1 \\
    -1           &  2+\epsilon^2  &  -1          &  ...       &  0          & 0 \\
    0            &  -1          &  2+\epsilon^2  &  ...       &  0          & 0 \\
    ...          &  ...         &  ...         &  ...       &  ...          & ... \\
    0            &  0           &  0           &  ...       &  2+\epsilon^2 & -1  \\
    -1           &  0           &  0           &  ...       &  -1         & 2+\epsilon^2 \\
  \end{array}
\right),\nn
\end{eqnarray}
with  $\epsilon=\xi/a$. The determinant of $A_n$ can then be calculated, and we get
\begin{eqnarray}
\det A_n &=& \sum\limits_{m=0}^{n-1}
\frac{\prod_{i=0}^{ m}(n^2-i^2)}{2^m (2m+1)!! (m+1)!} \epsilon^{2+2m} \nn \\
&& =
4\sinh^2 \left[n~ \mathrm{arcsinh}\frac{\epsilon}{2}\right].
\end{eqnarray}
Therefore, the action evaluated using the trial function~(\ref{eq-trial-fun})
can be written as
\begin{eqnarray}
S(E)&=& \int d x \left[  \frac{\hbar^2}{4m_0}( \phi_0^{\prime\prime}(x))^2 - \frac{1}{2} E {\phi}_0^2(x) \right] \nonumber \\
&-&\sum\limits_{n=1}^{\infty} \frac{(-1)^n}{2n} \frac{(I_0 \xi C)^n}{(  \sqrt{\pi} )^n}
\frac1{ \sinh\left[n ~\mathrm{arcsinh}\frac{\epsilon}{2}\right]}. \label{eq-50-1}
\end{eqnarray}
When $E\sim 1/{a^2} \ll E_\xi =1/{\left(2\xi^2\right)}$ we can approximate
$\sinh\left[n~ \mathrm{arcsinh}\frac{\epsilon}{2}\right] \approx n \epsilon /2$.
As a result the second line of Eq.~(\ref{eq-50-1}) is simplified to
$- \mathrm{Li}_2[- I_0 \xi C/\sqrt{\pi}]/\epsilon$.
This expression can be  also derived by applying the trial function method
directly to the action~(\ref{eq:action-av}). This means that
the action~(\ref{eq:action-av}) obtained
in the limit of uncorrelated speckle potential properly describes
the DOS in the presence of correlations for $E\ll E_\xi$. For these low
laying states the finite range correlations play no role because
the typical width of the wave functions $a$ is much larger than the
correlation length of disorder $\xi$.

\section{The Lifshitz tail for a particle in Gaussian uncorrelated disorder}
\label{appendix-C}

In order to illustrate the power of the GY and MT methods,
we reconsider here the problem of a particle in uncorrelated
Gaussian disorder. There existed a disagreement
in the literature on the preexponential factor in the asymptotic behavior
of the DOS in the tail of the band~\cite{Cardy78,Zittartz1966,yaida12}.
The two points that have not been sufficiently  discussed in these works
are the contribution of the continuous part of the spectrum to
the functional determinants and divergence of functional determinants
in $d\ge2$~\cite{Houghtont79}.
The replicated action of the system is given by~\cite{Cardy78}
\begin{eqnarray}\label{eq:action-av-cardy}
S_{\rm av}=\int d^{d}x &\left\{\frac{1}{2}(\nabla \bar{\phi}(x))^2
-\frac12 E\bar{\phi}^2(x) - \frac16 (\gamma/4)^2 (\bar{\phi}(x))^2
\right\} \nn \\
\end{eqnarray}
and we look for the asymptotic behavior of the DOS in the limit $E \to -\infty$.
The saddle point solution to the action~(\ref{eq:action-av-cardy}) has the form
\begin{eqnarray}
\bar{\phi}_{\rm cl} = \frac{(-E)^{1/2}}{\gamma}f(\sqrt{-E} x )~ \bar{n},
\end{eqnarray}
where $f$ satisfies the dimensionless equation of motion
\begin{eqnarray} \label{eq:schroedingerCardy}
\nabla^2 f-f=-\frac{f^3}{24}.
\end{eqnarray}
The action~(\ref{eq:action-av-cardy}) evaluated at the saddle point behaves
as  $S_{\rm cl}\sim (-E)^{2-d/2}/\gamma^2$.
By repeating the calculations~(\ref{eq-pref-2})-(\ref{eq:ratio})
we find the DOS tail
\begin{eqnarray} \label{eq-cardy12}
\nu(E) \sim (-E)^{1/2-d^2/4+3d/4} \sqrt{\frac {\det' M_\mathrm{T}}{\det' M_\mathrm{L}}}~
e^{-S_{\rm cl}},
\end{eqnarray}
where the regularized determinants of the fluctuation operators need to be calculated.
The transverse and longitudinal operators derived by expansion
around the saddle point solution have the form
\begin{eqnarray}\label{eq:MTCardy}
M_\mathrm{T} = - {\nabla}^2 - E\left[1-\frac{1}{24}f\left(\sqrt{-E} x\right)^2\right]
\end{eqnarray}
\begin{eqnarray}\label{eq:MLCardy}
M_\mathrm{L} =
- {\nabla}^2 - E\left[1-\frac{1}{8}f\left(\sqrt{-E} x\right)^2\right],
\end{eqnarray}
where the energy $E$ is assumed to be large and negative.

\subsection{Brute force method in $d=1$}\label{appendix-C1}

The one dimensional case is interesting for testing the
GY method because the ratio of determinants of
the operators~(\ref{eq:MTCardy}) and (\ref{eq:MLCardy}) can also be  calculated
directly  from the product
of their eigenvalues. The saddle point solution
of Eq.~(\ref{eq:schroedingerCardy}) is
\begin{equation}{\label{eq:classCardy}}
f(x)=4\sqrt{3}~{\rm sech}(x)
\end{equation}
and the operators~(\ref{eq:MTCardy})-(\ref{eq:MLCardy})
can be rewritten in the form of P\"oschl-Teller operators~\cite{Kleinert04,Dunne08}
\begin{eqnarray}\label{eq:Teller}
 {\cal M}_{m,j}= - \frac{\partial^2 }{\partial x^2}+
 |E|~\left[m^2 -j\left(j+1\right){{\rm sech^2}{\left(\sqrt{-E}~x\right)}}\right], \nn \\
\end{eqnarray}
where $j$ takes integer values. The operators~(\ref{eq:MTCardy}) and
(\ref{eq:MLCardy}) correspond
to the case $M_T={\cal{M}}_{1,1}$ and $M_L={\cal{M}}_{1,2}$.
The spectrum of the P{\"o}schl-Teller operator ${\cal M}_{m,j}$
contains a discrete part which have $j$ bound states
$\varepsilon_{\ell} =|E| \left(m^2-\ell^2\right)$ ($\ell=1,...,j$) and the continuous
part $\varepsilon(k)=|E| \left(k^2+m^2\right)$ with the density of states
\begin{eqnarray}\label{eq:densM}
\nu_j\left(k\right)=\frac{L}{2\pi}-\frac{1}{\pi}\sum_{\ell=1}^j{\frac{\ell}
{\ell^2+k^2}},
\end{eqnarray}
which has been regularized by putting the system in a box of size $L$.
This yields
\begin{eqnarray}\label{eq:logdetM}
&&\!\! \log \frac{\det{{\cal{M}}_{m,j}}}{\det{{\cal{M}}_{m,0}}}=
 \sum_{\ell=1}^j{\log{\left[|E| \left(m^2-\ell^2\right)\right]}} \nn \\
&& \ \ \ \ +\int^{\infty}_{-\infty} d
 k ~\nu_j\left(k\right)\log{\left[|E|~\left(k^2+m^2\right)\right]},
\end{eqnarray}
where the free particle contribution $L/2 \pi$ has been canceled.
From this formula the functional determinants of
the operators~(\ref{eq:MTCardy})-(\ref{eq:MLCardy})
can be determined straightforwardly. The operator $M_T={\cal{M}}_{1,1}$ has just
one discrete zero eigenvalue. Excluding this zero mode, the only contribution
comes from
the continuous spectrum given by the second line in Eq.~(\ref{eq:logdetM}),
\begin{eqnarray}\label{eq:logdetMTcardy}
\frac{\det{{{'{M}}}_{T}}}{\det M_{\rm free}}=\frac{1}{4 |E|},
\end{eqnarray}
where $M_{\rm free}:={\cal{M}}_{m,0}$.
The operator $M_L={{M}}_{1,2}$ has two discrete eigenvalues:
a negative eigenvalue ($\ell=2$) giving a
finite imaginary part of the Green function~(\ref{eq:dens}),
and a zero eigenvalue ($\ell={1}$). The negative mode and the continuous spectrum
contribute with
\begin{eqnarray}\label{eq:logdetMLcardy}
\frac{\det{{{'{M}}}_{L}}}{\det M_{\rm free}}=-\frac{1}{12|E|}.
\end{eqnarray}
We obtain
\begin{eqnarray}\label{eq:ratiocardy}
\lim_{N\rightarrow 0}\det{'M_L}^{-1/2} \cdot \det{'M_T}^{-(N-1)/2}=i~\sqrt{3},
\end{eqnarray}
which is independent of $E$.

\subsection{Regularized Gel'fand-Yaglom formula}\label{appendix-C2}

The same result can also be obtained  using  the GY method~(\ref{eq:GY}).
The solution $u_{m,j}$ of the corresponding Cauchy
problem~(\ref{eq:eigen_probl-1})-(\ref{eq:eigen_probl})
for the  P\"oschl-Teller operators~${\cal M}_{m,j}$ (\ref{eq:Teller})
can be found analytically. For the sake of
compactness, we show here the formula for $|E|=1$,
\begin{eqnarray}\label{eq:GYcardysol}\hspace{-2.5cm}
&&u_{m,j}(L)=\frac{1}{\left(1+j-m\right)} \nn \\
&& \times \frac{-P^m_j
\left(  Y \right)Q^m_j\left(-  Y\right)+P^m_j
\left(-  Y\right)Q^m_j\left(  Y\right)}{P^m_{j+1}
\left(-  Y\right)Q^m_j\left(-  Y\right)-P^m_j\left(-  Y\right)
Q^m_{j+1}\left(-  Y\right)}, \nonumber \\
\end{eqnarray}
where we defined $Y:=\tanh L$ and $P^m_j(x)$ and $Q^m_j(x)$ are the Legendre
functions.
The solution~(\ref{eq:GYcardysol}) for the free operator
${\cal M}_{\rm free}={\cal M}_{m,0}$ reduces to
\begin{eqnarray}\label{eq:GYcardysolfree}
u_{m,0}(L)=\frac{\sinh{\left[2 m L\right]}}{m}.
\end{eqnarray}
The ratio of the two determinants  is then given by
\begin{eqnarray}\label{eq:GYbis}
\frac{\det{\cal M}_{m,j}}{\det{\cal M}_{m,0}}= \lim \limits_{L\to \infty}
\frac{u_{m,j}(L)}{u_{m,0}(L)}.
\end{eqnarray}
In order to exclude the zero modes, we apply a shift of the mass
$m \rightarrow \sqrt{m^2+\delta m^2}$ that gives
\begin{eqnarray}\label{eq:renbis}
\frac{\det\left({\cal M}_{m,j}+\delta m^2\right)}{\det\left({\cal M}_{m,0}+\delta m^2\right)}\sim \delta m^2
\frac{\det{'\cal M}_{m,j}}{\det{\cal M}_{{m,0}}}, \hspace{2mm}
\delta m^2\rightarrow 0. \ \nn
\end{eqnarray}
Restoring the dependence on $E$ we
obtain
\begin{eqnarray}\label{eq:detEnergydep}
\frac{\det{'\cal M}_{1,j}}{\det{\cal M}_{{1,0}}}=(-1)^{j+1}\frac{1}{2j(j+1)|E|}
\end{eqnarray}
in agreement with Eq.~(\ref{eq:logdetMTcardy})-(\ref{eq:logdetMLcardy}) for $j=1$ and $j=2$ respectively.

\subsection{McKane-Tarlie formula}\label{appendix-C3}

We now apply the MT method. First, we need the
zero modes of the operators $M_T={\cal M}_{1,1}$ and $M_L={\cal M}_{1,2}$. They are
given by $v_{1,1}(x)=|E| f (\sqrt{-E} x )$ and
$v_{1,2}(x)=|E|^{3/2} f' (\sqrt{-E} x )$, respectively.
Then, according to Eq.~(\ref{eq:limitdet}), the ratio of the
determinants with excluded zero modes is given by
\begin{eqnarray} \label{eq:limitdet2}
\frac{\det'M_T}{\det' M_L}=\lim_{L\rightarrow\infty}
\frac{\left\langle v_{1,1}| v_{1,1}\right\rangle}{\left\langle
v_{1,2}|v_{1,2}\right\rangle}
\frac{v_{1,2}^{\prime}(-L)
v_{1,2}^{\prime}(L)}{v_{1,1}^{\prime}(-L)
v_{1,1}^{\prime}(L)}=-3, \nn \\
\end{eqnarray}
where we used
\begin{eqnarray}
&&\left\langle v_{1,1}| v_{1,1}\right\rangle= 2 |E|^{3/2}, \\
&&\left\langle v_{1,2}| v_{1,2}\right\rangle= \frac23 |E|^{5/2},\\
&&\frac{v_{1,2}^{\prime}(-\infty)}{v_{1,1}^{\prime}(-\infty)}
= - \frac{v_{1,2}^{\prime}(\infty)}{v_{1,1}^{\prime}(\infty)}=\sqrt{-E}.
\end{eqnarray}

\subsection{Gel'fand-Yaglom method generalized to radial operators for $d>1$}
\label{sec-GY-higher-dimensions}

The radial parts of the eigenfunctions of the
operators~(\ref{eq:MTCardy}) and (\ref{eq:MLCardy}) satisfy
Eq.~(\ref{eq-rad-operators-1})
with the mass $m=|E|$ and the potentials
\begin{eqnarray}
U_{T}(r)=- \frac{|E|}{24}f\left(\sqrt{-E} r\right)^2 ,
\label{eq-potentials-9-1} \\
U_{L}(r)= -\frac{|E|}{8}f\left(\sqrt{-E} r\right)^2.
\label{eq-potentials-10-1}
\end{eqnarray}
Scaling analysis shows that the solutions of the corresponding
Cauchy problems (\ref{eq:bound-cond-1}) and (\ref{eq:bound-cond-2})   have the form
\begin{eqnarray}
u^{(\ell)}&=&|E|g^{(\ell)}(|E|^{1/2}r), \\
u^{(\ell)}_{\rm free}&=&|E|g^{(\ell)}_{\rm free}(|E|^{1/2}r).
\end{eqnarray}
There is no zero modes for $\ell>1$ and we can apply the GY formula~(\ref{eq:G-Y3d}).
The ratio of the partial determinants for $\ell>1$ is given by
\begin{equation}
\lim_{R\to \infty} g^{(\ell)}(|E|^{1/2}R)/g^{(\ell)}_{\rm free}(|E|^{1/2}R),
\end{equation}
which does not depend on $E$. Thus, all the ratios of the partial determinants
with $\ell>1$ do not contribute to the $E$ dependance of the full ratio
of the functional determinants.

The operators ${M}_{T}^{(0)}$ and ${M}_{L}^{(1)}$ have a zero eigenvalue so that
to exclude it we apply the MT method.
The corresponding ratios of the partial determinants can be calculated using
Eq.~(\ref{eq:limitdet-3}).
The scaling behavior of the zero mode eigenfunction $v_0(r)$  is again given by
 \begin{eqnarray}
&& v_0(r)=|E|g_0(|E|^{1/2}r).
\end{eqnarray}
This yields
\begin{eqnarray}
&&\left\langle v_0 |v_0 \right\rangle \sim |E|^{3/2}, \\
&& \lim_{R\to \infty} v_0'(R) u_{\rm free} (R) = |E|^{5/2}
\lim_{R\to \infty} g_0^{\prime}(R)g_{\rm free}(R), \ \ \ \
\end{eqnarray}
where the last limit is expected to be finite.
Thus, the logarithms of the determinant ratios are given (up to an energy
independent constants)  by
\begin{eqnarray}\label{eq:MT3d}
\ln{\frac{\det{'M}_{T}}{\det{M}_{\rm free}}}
={\rm deg}\left(0;d \right)~\ln{\frac{\det'{M}_{T}^{(0)}}{\det{M}_{\rm free}^{(0)}}}
\sim \ln{|E|^{-1}} \nonumber \\
\end{eqnarray}
and
\begin{eqnarray}\label{eq:ML3d}
\ln{\frac{\det{'M}_{L}}{\det{M}_{\rm free}}}
={\rm deg}\left(1;d \right)~\ln{\frac{\det'{M}_{L}^{(1)}}{\det{M}_{\rm free}^{(1)}}}
\sim d\ln{|E|^{-1}}. \nonumber \\
\end{eqnarray}
Above we assumed that the ratios of the determinants are finite and
the sum over $\ell$ is converging.
However, we know that this sum diverges for $d=2$ and $d=3$. We have to
subtract from each ratio of the partial determinants
the term resulting from the partial wave decomposition
of the diverging diagram $A_1$.  After that the regularized
diagram $A_1$ has to be added to the action as shown in Eq.~(\ref{eq:sum-finite}).
The terms needed to be subtracted are given by Eq.(\ref{jost-expansion}) and read
\begin{eqnarray} \label{eq-subtr-1}
\int\limits_0^{\infty}dr\, r\, U_{T,L}(r)
K_{\nu} (|E|^{1/2} r) I_{\nu} (|E|^{1/2} r)
\end{eqnarray}
where  $\nu\equiv l+\frac{d}{2}-1$ and $ U_{T,L}$ are given by
Eqs.~(\ref{eq-potentials-9-1}) and (\ref{eq-potentials-10-1}). It is easy to see
that the terms (\ref{eq-subtr-1}) do not depend on $E$.
The bare diagram $A_1$ can be written as
\begin{eqnarray} \label{eq-diag-A1}
A_1^{T,L} = \int d^dx\, U_{T,L}(|x|) \lim\limits_{y\to x} C_0(x-y),
\end{eqnarray}
where $C_0(x)$ is given by Eq.~(\ref{eq-G-1}).
The expression~(\ref{eq-diag-A1})
diverges and has to be regularized, e.g. by the UV cutoff $\Lambda$ as follows
\begin{eqnarray} \label{eq-A11}
A^{T,L}_1=\int_{|q|<\Lambda} \frac1{q^2+m^2}\,  \int d^d x\,  U_{T,L}(|x|),
\end{eqnarray}
where the last integral  behaves as $(-E)^{1-d/2}$.
Combining Eqs.~(\ref{eq:MT3d})-(\ref{eq:ML3d}) and (\ref{eq-A11}) we obtain
\begin{eqnarray}\label{eq:cardy3d}
\sqrt{\frac{\det{'M}_{T}}{\det{'M}_{L}}}\sim i |E|^{\frac{d-1}{2}} e^{(A^T_1-A_1^L)/2},
\end{eqnarray}
where the factor $i$ comes from the negative eigenvalue of the
partial operator ${M}^{(0)}_{L}$.
Inserting Eq.~(\ref{eq:cardy3d}) into Eq.~(\ref{eq-cardy12})
gives the DOS tail for $E \to -\infty$. Additionally to the
term $(-E)^{2-d/2}$ resulting from the action evaluated at the saddle point
the exponential now contains the term $(-E)^{1-d/2}$ with a non-universal
coefficient which depends on the UV cutoff. Fortunately, we already
know how to renormalize the field
theory~(\ref{eq:action-av-cardy}) which is nothing but $\phi^4$ theory.
It easy to recognize in the first integral
in Eq.~(\ref{eq-A11}) the one-loop diagram which shifts the mass $m^2=|E|$. The
diagram (\ref{eq-A11}) is compensated by the counterterm coming from the mass shift.
Thus, in terms of the renormalized energy $E_R = E - E_0$, where $E_0$ is some
non-universal energy scale depending of the UV cutoff, the DOS tail has the form
\begin{eqnarray}
\nu( E_R ) \sim (-E_R)^{d(5-d)/4} e^{-\mathrm{const} (-E_R)^{2-d/2}/\gamma^2}.
\end{eqnarray}

%\section*{References}

\end{document}